\documentclass[epjCONF,final]{svjour} 
\usepackage{graphicx}
\usepackage{subfig}
\usepackage[dvips]{epsfig}
\usepackage{color}
\usepackage{epstopdf}
\usepackage{url}
\usepackage[varg]{txfonts} 
\usepackage[latin1]{inputenc}

\session-title{EPJ Web of Conferences}
\begin{document}
\title{Simulation of magnetic active polymers for versatile microfluidic devices\footnote{Published in \textit{EPJ Web of Conferences}, DOI:http://dx.doi.org/10.1051/epjconf/20134002001}}
\author{Markus Gusenbauer\inst{1}\fnmsep\thanks{\email{markus.gusenbauer@fhstp.ac.at}} \and Harald \"Ozelt\inst{1} \and Johann Fischbacher\inst{1} 
\and Franz Reichel\inst{1} \and Lukas Exl\inst{1} \and Simon Bance\inst{1} \and Nadezhda Kataeva\inst{2} \and Claudia Binder\inst{2} \and Hubert Br\"uckl\inst{2} \and Thomas Schrefl\inst{1} }
\institute{St. Poelten University of Applied Sciences, St. Poelten, Austria \and AIT Austrian Institute of Technology, Vienna, Austria }
\abstract{
We propose to use a compound of magnetic nanoparticles (20--100\,nm) embedded in a flexible polymer (Polydimethylsiloxane PDMS) to filter circulating tumor cells (CTCs). The analysis of CTCs is an emerging tool for cancer biology research and clinical cancer management including the detection, diagnosis and monitoring of cancer. The combination of experiments and simulations lead to a versatile microfluidic lab-on-chip device. Simulations are essential to understand the influence of the embedded nanoparticles in the elastic PDMS when applying a magnetic gradient field. It combines finite element calculations of the polymer, magnetic simulations of the embedded nanoparticles and the fluid dynamic calculations of blood plasma and blood cells.  With the use of magnetic active polymers a wide range of tunable microfluidic structures can be created. The method can help to increase the yield of needed isolated CTCs.
} 

\maketitle

\section{Introduction}
\label{sec:intro}
Circulating tumor cells (CTCs) detach from a tumor and can remain in the blood even after the tumor is removed. Their presence increases the chance of new tumors developing. It is important to monitor the number of CTCs in the blood but their low concentration when compared to normal blood cells makes this difficult (one CTC per 5--10 million blood cells). A new and versatile method to isolate every single CTC is required. Due to the low concentration of CTCs a capture failure could lead to a wrong diagnosis and bad decisions for treatment.

The most common approaches for capturing CTCs are mechanical based on the size isolation \cite{lu_parylene_2010}, using antibodies based on the affinity mechanisms \cite{bell_isolation_2007}, or using a density separation method \cite{gertler_detection_2003}. Recent technologies try to combine the most promising methods for better filter results. In our preliminary work we show how the CTC yield could be increased with a tunable magnetic bead structure \cite{gusenbauer_self-organizing_2012}.

Our proposed chip technology may offer the possibility to combine affinity and size capturing through changing the cross section of a microfluidic channel with an external magnetic gradient field (Section \ref{subsec:tuneChannel}). The channel walls are covered with a tumor specific antibody. Because of affinity mechanisms the CTCs are captured when getting into contact with the walls. Turbulences in the fluid increase this possibility. Different structures like herringbones \cite{stott_isolation_2010} can be used to mix the laminar fluid lines. The height of the microstructure can be changed with the external field creating an additional mechanical filter for CTCs.

To create a computer model of the tunable device a cantilever beam experiment was established for validation (Section \ref{subsec:validationPolymer}). Computer simulations with the finite element software tools Elmer \cite{elmer} and Abaqus \cite{abaqus} in combination with experiments show the advantages of this new approach. 

\section{Model of magnetic active polymer}
\label{sec:modelPolymer}

A flexible polymer (Polydimethylsiloxane PDMS) can be manipulated with a magnetic field by embedding nanoparticles \cite{nanofer}. This particles consists of an iron core (Fe(0)) and a ferroxide-carbon shell ($\mathrm{Fe_3O_4}$--C). They have a crystallite size of around 20--100\,nm. In order to calculate the behaviour of the magnetic active polymer with an applied magnetic field the finite element simulation environments Elmer (free \cite{elmer}) and Abaqus (commercial \cite{abaqus}) are used. Material parameters for the polymer and the nanoparticles are given in table \ref{tab:paramPolymer}. A 3-dimensional mesh of the polymeric model is the basis for the finite element solvers of the software tools. 

\begin{table}
\begin{tabular}{lrl}
  \hline\noalign{\smallskip}
  & \textbf{value} & \textbf{unit} \\
  \noalign{\smallskip}\hline\noalign{\smallskip}
  \textbf{polymer} & &\\
  density & $1210$ & $\mathrm{kg\,m^{-3}}$\\
  Young's modulus & $2$ & $\mathrm{MPa}$\\
  Poisson's ratio & $0.45$ & \\
  \noalign{\smallskip}\hline\noalign{\smallskip}
  \textbf{nanoparticles} & &\\
  concentration in polymer & 1--5 & $\mathrm{vol\%}$\\
  radius& 10--50 & $\mathrm{nm}$ \\
  mean radius (Scherrer formula)& 25 & $\mathrm{nm}$ \\
  spec. magnetic moment saturated & $178$ & $\mathrm{A\,m^2\,kg^{-1}}$\\
  density & $7870$ & $\mathrm{kg\,m^{-3}}$\\
  \noalign{\smallskip}\hline\noalign{\smallskip}
  \textbf{magnetic field} & &\\
  current& $1.5$ & $\mathrm{A}$ \\
  gradient & $1.78$ & $\mathrm{dB\ dx^{-1}}$\\
  \noalign{\smallskip}\hline
\end{tabular}
 \caption{Parameters of polymer, nanoparticles and external magnetic field. More details on the nanoparticles are given on the developers homepage \cite{nanofer}}
\label{tab:paramPolymer}
\end{table}

The magnetic force on a single embedded nanoparticle is given by the negative gradient of the energy of the magnetic dipole moment $\vec m$ in the field $\vec B$.

\begin{equation}
\vec F_g=\nabla(\vec m \cdot \vec B)
\label{eqn:Fg}
\end{equation}

In Elmer the magnetic force is implemented as a body force $\vec F_b$ (force per volume) acting on the polymer. With the known concentration of nanoparticles in the PDMS this body force can be extracted (Eqn. \ref{eqn:Fb}). It is the number of beads $N$ in a control volume $V$ of the PDMS times the gradient force $\vec F_g$ on a single particle divided by the control volume $V$. 

\begin{equation}
\vec F_b=\frac{N\ \vec F_g}{V} 
\label{eqn:Fb}
\end{equation}

The crucial part of this equation is the amount of beads $N$ (Eqn. \ref{eqn:N}). It depends on the density of the nanoparticle powder $d$ (sphere pack problem) which can only be approximated, the known concentration in the polymer $c$ and the control volume $V$ over the particle volume $V_{bead}$. 

\begin{equation}
N=\frac{V\ c\ d}{V_{bead}} 
\label{eqn:N}
\end{equation}

After applying the force a finite element solver calculates the deformation of the object. Results of the validation and further applications of the magnetic active polymer are discussed in section \ref{sec:results}.


\section{Results}
\label{sec:results}

\subsection{Validation of the polymeric model}
\label{subsec:validationPolymer}

\begin{figure}
\centering
  \subfloat[]{\epsfig{file=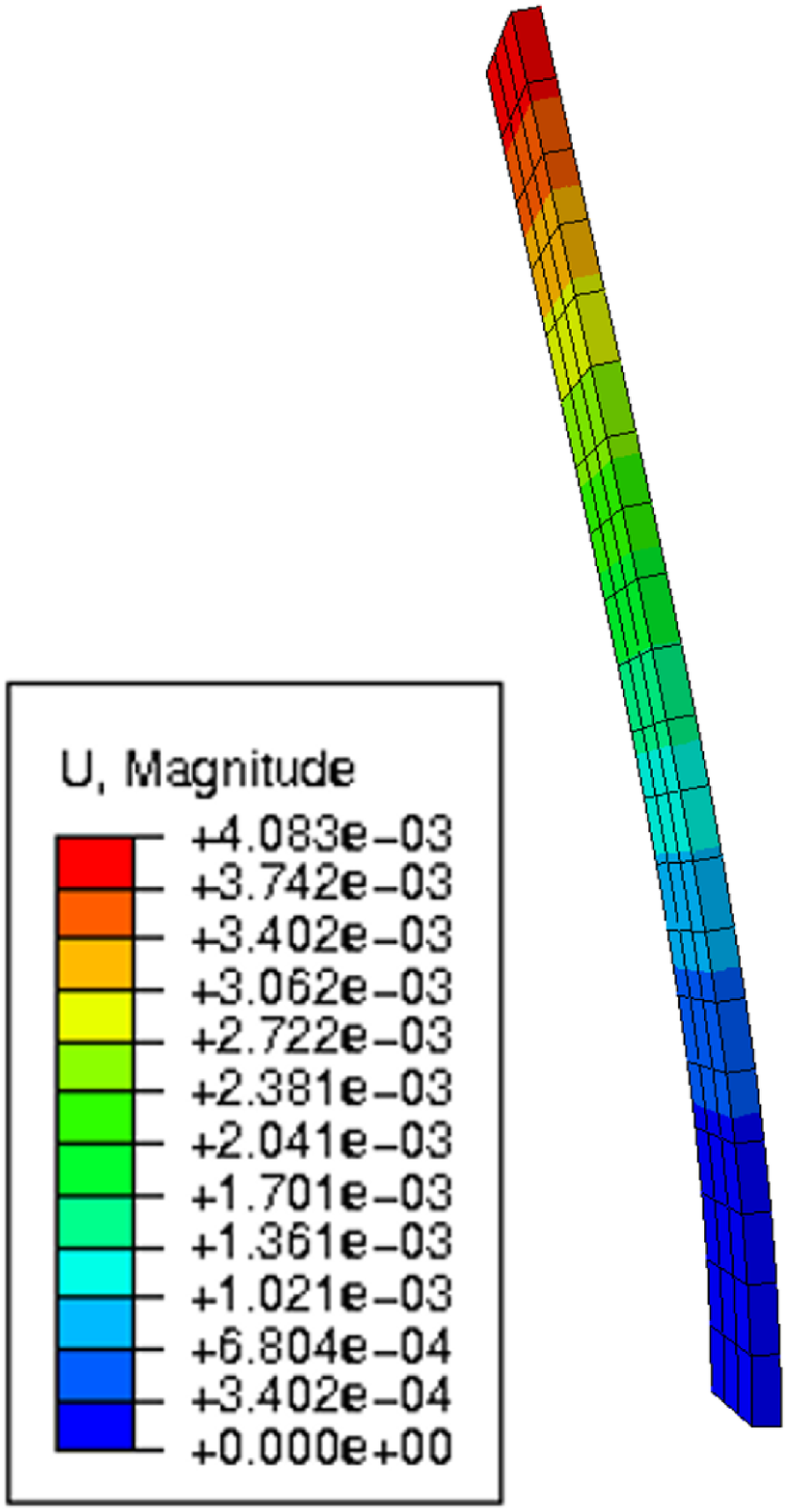,width=0.2\textwidth}\label{fig:beamAbaqus}}\qquad
  \subfloat[]{\epsfig{file=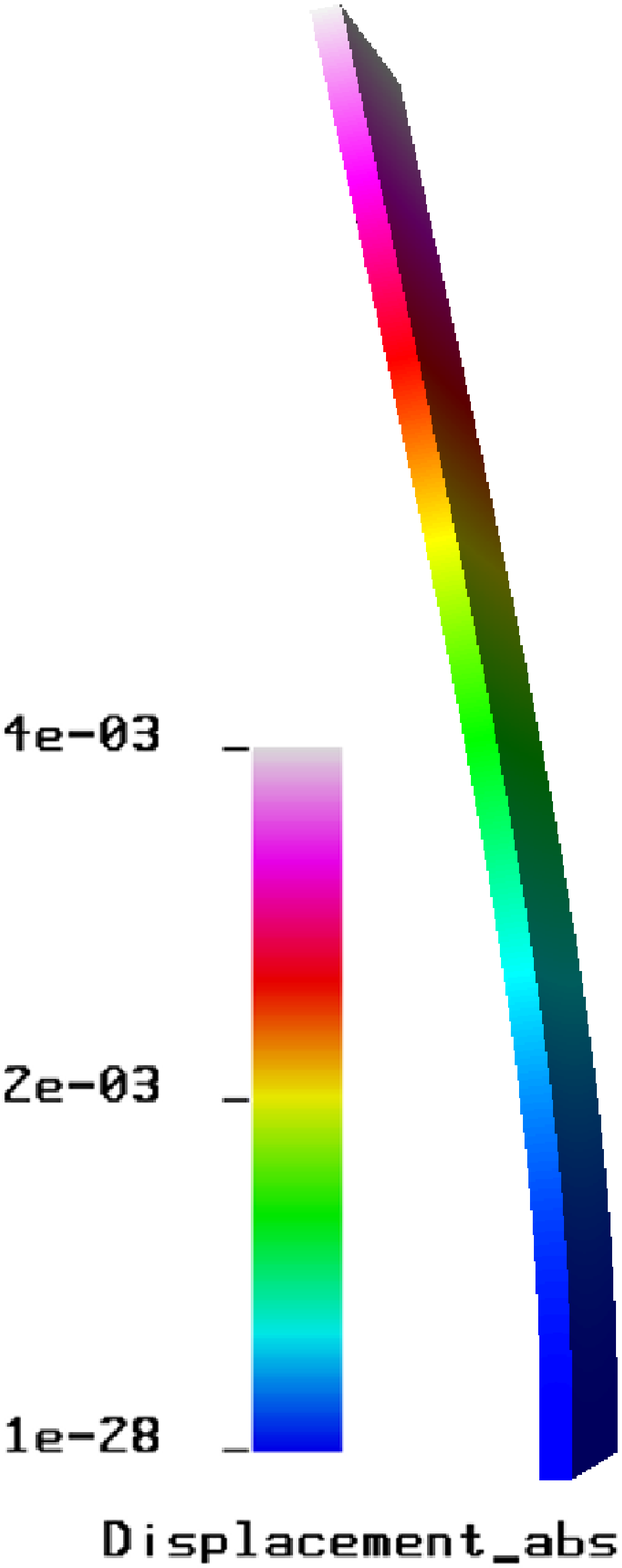,width=0.165\textwidth}\label{fig:beamElmer}}\qquad
  \subfloat[]{\epsfig{file=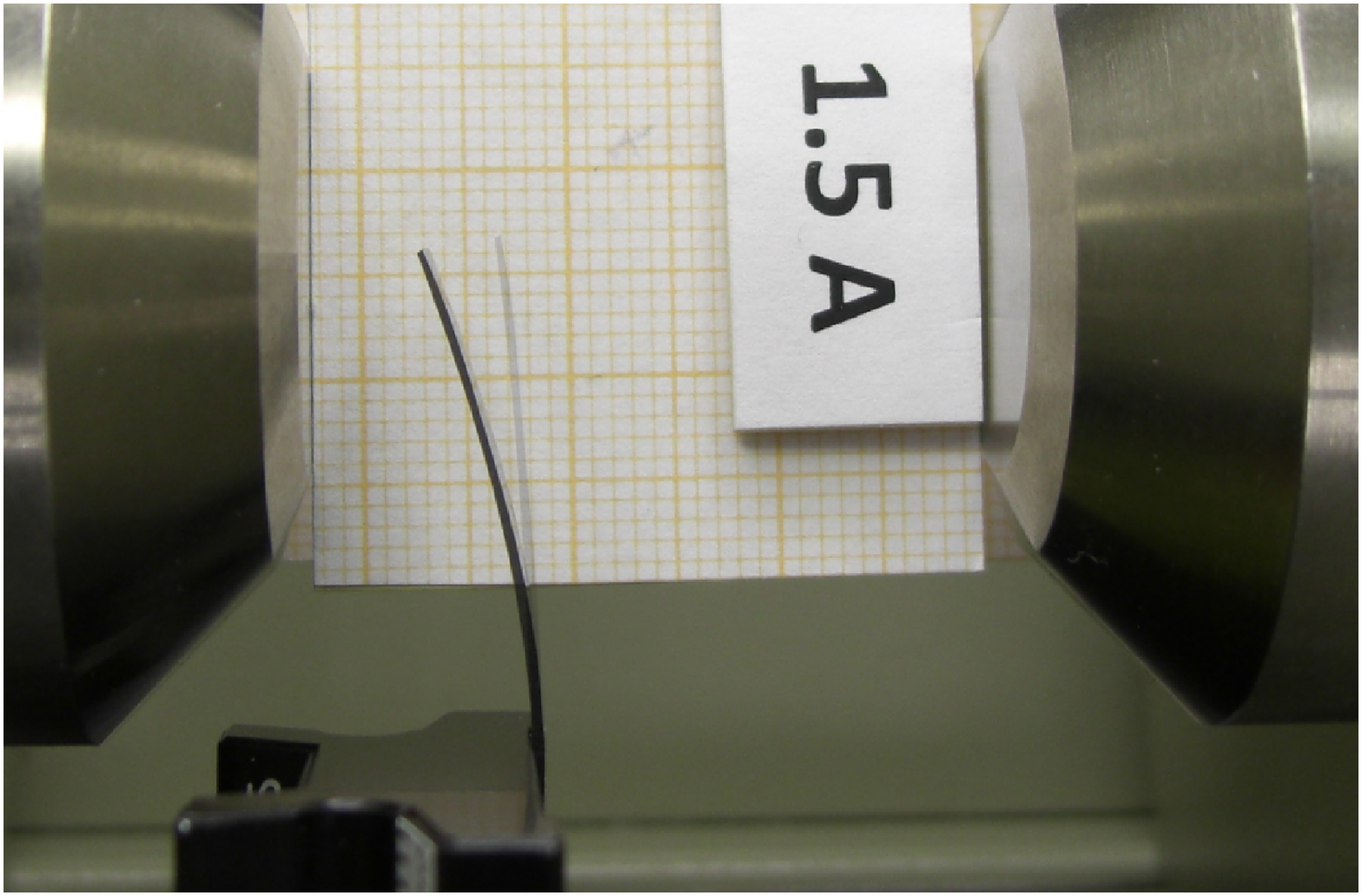,width=0.44\textwidth}\label{fig:beamExperiment}}\qquad
\caption{
    A body force $F_b$ (Eqn. \ref{eqn:Fb}) acts on a simulated elastic beam in Abaqus \protect\subref{fig:beamAbaqus} ($F_b = \mathrm{40\,kN\;m^{-3}}$) and in Elmer \protect\subref{fig:beamElmer} ($F_b = \mathrm{50\,kN\;m^{-3}}$). Both simulations cause an absolute deformation of $4\,mm$. This body force is a function of the magnetic gradient force $F_g$ (Eqn. \protect\ref{eqn:Fg}), which bends a cantilever beam in an experiment \protect\subref{fig:beamExperiment} (overlay of semi-transparent and non-transparent beam - before and after applying a current of $1.5\,A$, which creates a gradient field of $1.78$ $\mathrm{dB\ dx^{-1}}$ ).
}
\label{fig:beam}
  \end{figure}

Adjustment and validation of the simulated polymeric model and the real compound of nanoparticles and PDMS was done by cantilever beam experiments (Fig. \ref{fig:beam}c). A polymer stripe is fixed on one side. A magnetic gradient field creates a bending which can be compared with the simulations.

We have two possibilities to receive the body force $\vec F_b$. On the one hand we can calculate it with Eqn. \ref{eqn:Fb}. On the other hand we can vary the body force $\vec F_b$ in the simulations until the bending is the same as in the experiment. Both methods should deliver the same or a similar result.

The calculated body force $\vec F_b$ for nanoparticles with a mean radius of $\mathrm{25\,nm}$, a sphere pack density of $50\,\%$ (sphere pack problem, as mentioned in Section \ref{sec:modelPolymer}) and concentration of $\mathrm{3.5 \,vol\%}$ has a value of $\mathrm{43.79\,kN\;m^{-3}}$ (Eqn. \ref{eqn:Fb}). Due to an unknown distribution of the particles radii, the sphere pack density of the powder and the mixture of particles in the PDMS this input value is only an approximation for the simulations. The bending of the cantilever beam of $4\,mm$ (Fig. \ref{fig:beam}c) could be recreated in simulations with an input body force $\vec F_b$ of $\mathrm{40\,kN\;m^{-3}}$ (Abaqus Fig. \ref{fig:beam}a) and $\mathrm{50\,kN\;m^{-3}}$ (Elmer Fig. \ref{fig:beam}b).

Both values of $\vec F_b$ in Abaqus and Elmer, to cause similar deformation of the beam, are quite close to $\mathrm{43.79\,kN\;m^{-3}}$. Differences of the simulation tools, although they are using the same mesh and material parameters can be explained because of different finite element solvers.

\subsection{Tunable microfluidic channel}
\label{subsec:tuneChannel}

In this section the validated polymeric model is used to create a tunable microfluidic channel. For further investigations we prefer Elmer. Abaqus has a limitation of nodes, which are the corners of the tetrahedrons of the 3-dimensional mesh (only student version available). It prevents us of running more complex simulations.

\begin{figure}
\begin{center}
 \includegraphics[scale=.35]{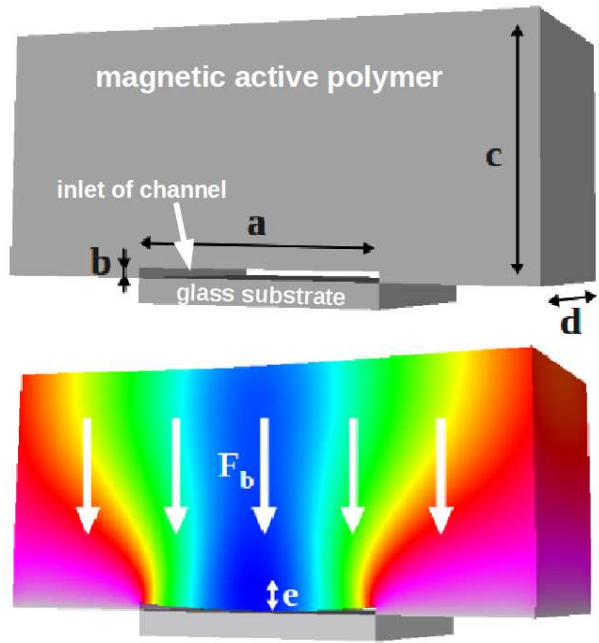}
\end{center}
 \caption{A Microfluidic channel is cut out of a magnetic active polymer block (top). An applied magnetic gradient field $F_g$ (Eqn. \protect\ref{eqn:Fg}), 
which in simulations act as a body force $F_b$ (Eqn. \protect\ref{eqn:Fb}), decreases the channel cross section (bottom).}
 \label{fig:channelSim}
\end{figure}

\begin{figure}
\begin{center}
 \includegraphics[scale=.29]{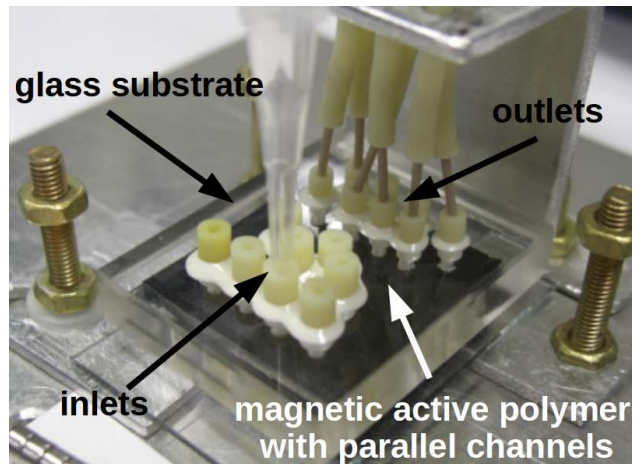}
\end{center}
 \caption{Experimental setup of tunable $\mu$-fluidic chip device. Magnetic active polymer (black) is placed on a glass substrate. Several in- and outlets indicating parallel fluidic channels.}
  \label{fig:channelSetup}
\end{figure}

The magnetic active polymer is the main part of the lab-on-chip device. A microfluidic channel is cut out of the PDMS with a large width to height ratio $a\gg b$ (Fig. \ref{fig:channelSim}). This block is placed on a glass base plate for better visualization (Fig. \ref{fig:channelSetup}). The polymer is black because of the enclosed softmagnetic particles, which prevents the use of a microscope from top- or side-view of the channel. At the start and at the end of the channel a fluidic in- and outlet is placed. Now a controlled fluid stream can pass the magnetic active polymer.

\begin{figure}
\begin{center}
 \includegraphics[scale=.3]{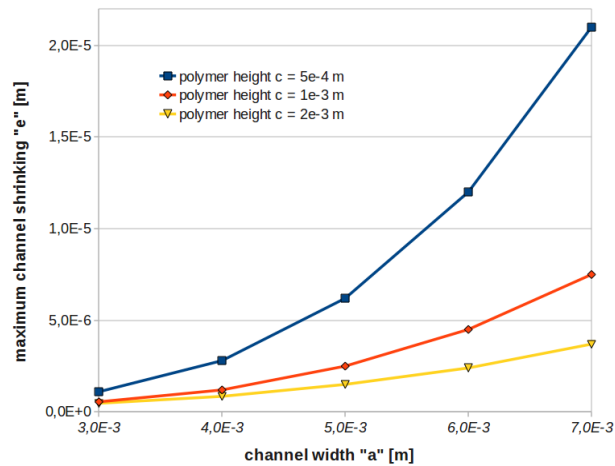}
\end{center}
 \caption{Influence of the geometry of the polymer (Fig. \protect\ref{fig:channelSim}) on the bending of the channel due to a magnetic gradient field. Most deformation can be established with a wide channel a and a small polymer height c.}
  \label{fig:geometry}
\end{figure}

Only by applying a magnetic gradient field the channel cross section can be changed. Similar to the cantilever beam experiment in section \ref{sec:modelPolymer} the magnetic gradient force bends the channel ceiling. The geometry of the polymer block is essential for the bending behavior (Fig. \ref{fig:channelSim}). The largest deformation occurs for a wide channel a and a small polymer height c (Fig. \ref{fig:geometry}). With the cantilever beam experiments in section \ref{subsec:validationPolymer} we have seen that the relative bending is little compared to the length of the beam. Therefore a magnetic gradient field does not create a considerable deformation of a narrow channel. The channel length d is not important for the bending behavior. In the direction of the fluid flow the magnetic field is assumed to be homogeneous. No gradient field and therefore no magnetic force in the flow direction is applied to the embedded particles in the PDMS. The channel height b does not influence the bending too, 
provided that the width to height ratio is $a\gg b$.


\section{Conclusion and future work}
\label{sec:conclusion}

Versatile filters are important to get a high probability of catching cancer cells. A magnetic active polymer is used to create a tunable microfluidic channel. Only by applying an external magnetic gradient field the channel cross section can be modified. This knowledge can help to increase the yield of CTC filter devices.

Several approaches for capturing single CTCs like the herringbone-chip \cite{stott_isolation_2010} can be improved. Stott et al. showed that turbulences in the blood flow are important to capture them. He created a microfluidic chip with a herringbone like wall structure. This bones are mixing the fluid lines and the probability of blood cells getting into contact with the channel walls is increasing. Without mixing the flow would be highly laminar. The walls are covered with a tumor specific antibody which captures the CTCs using affinity mechanisms. With the magnetic active polymer this method can be improved. In order to capture different CTCs (varying in size) and to further increase the turbulences of the blood flow the channel height can be changed with an external magnetic gradient field. Other advantages could be that the tunable channel dissolves cell clumps or disperse other blocking like unwanted air bubbles just by changing the height.

\begin{figure}
   \centering
      \subfloat[]{\epsfig{file=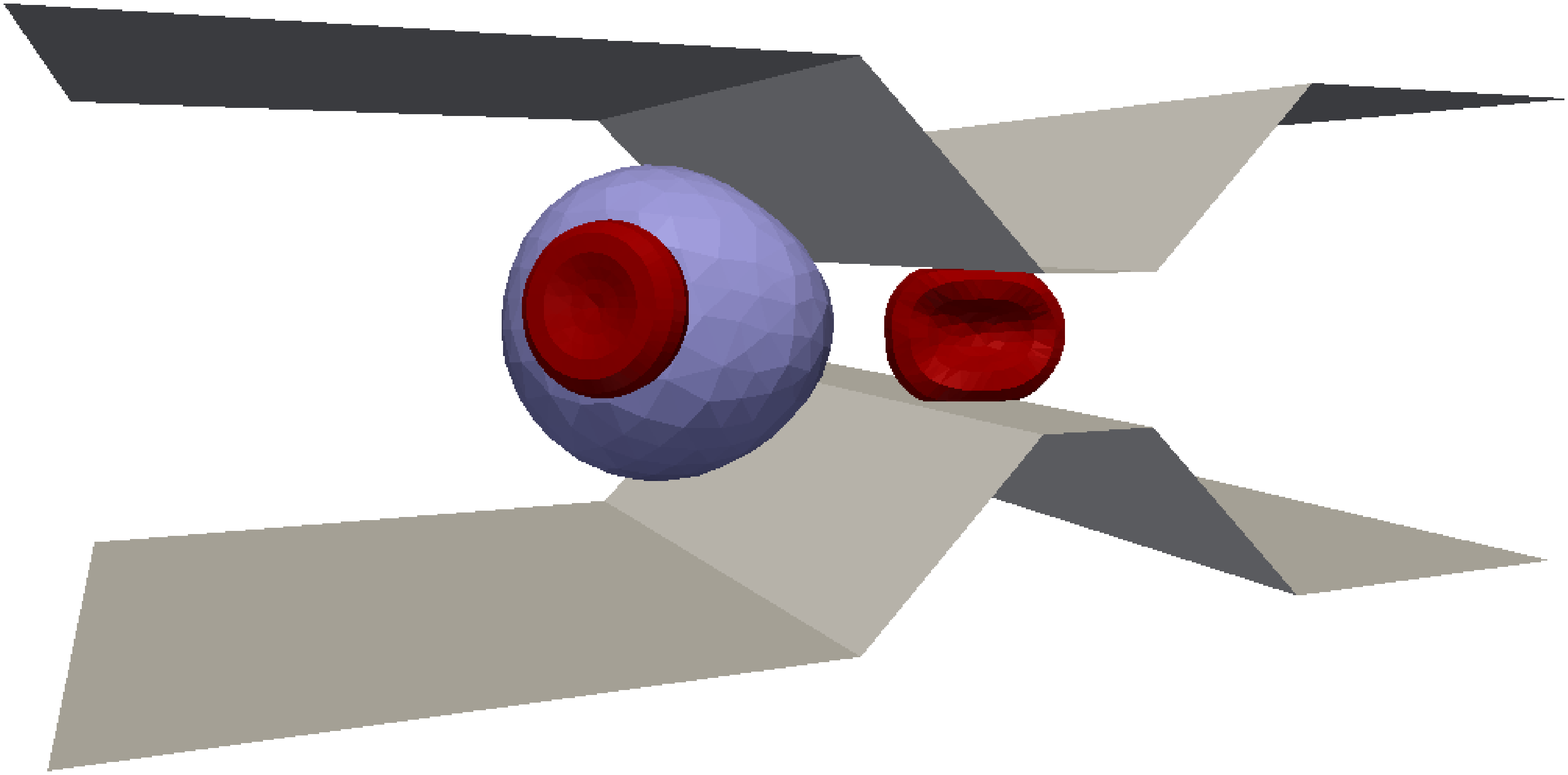,width=0.3\textwidth}\label{fig:mingap}}\qquad
      \subfloat[]{\epsfig{file=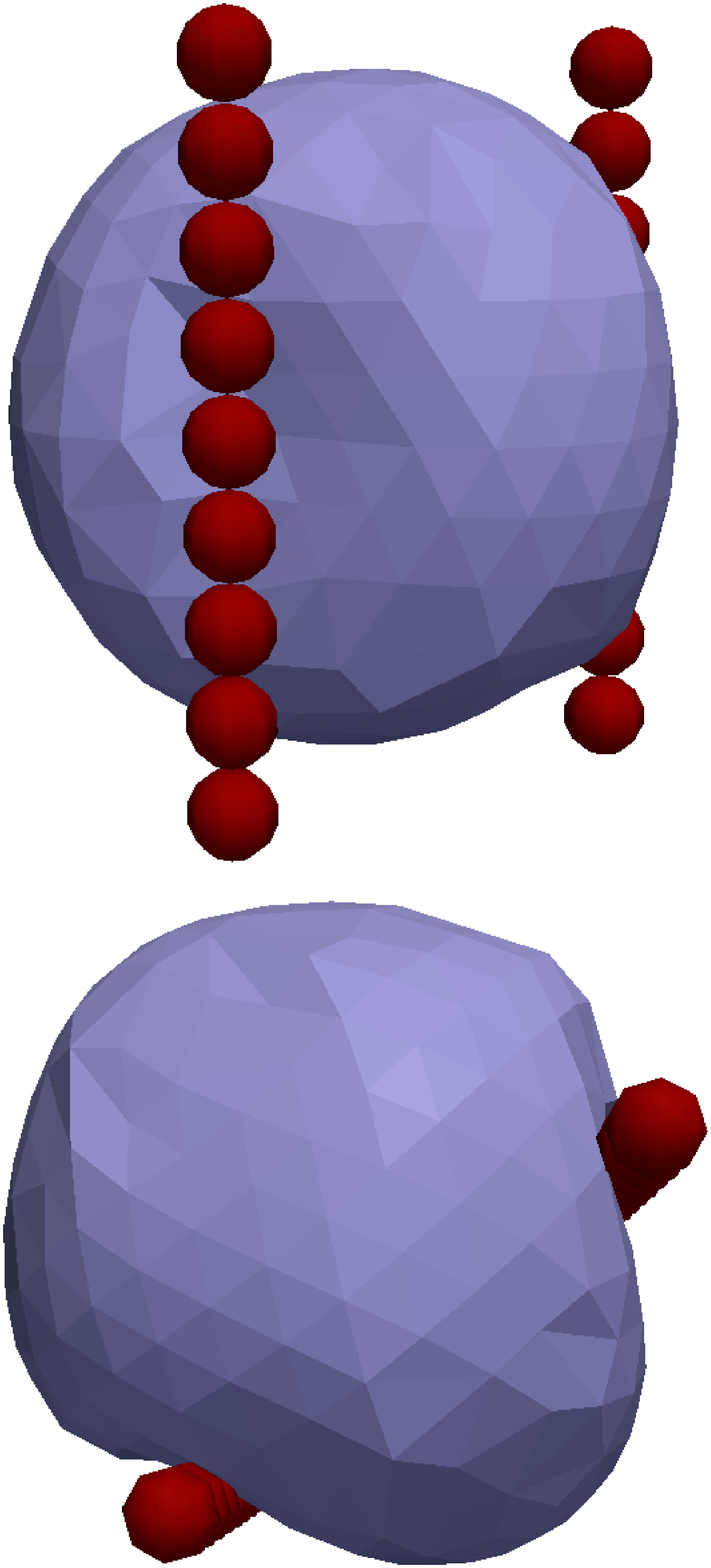,width=0.07\textwidth}\label{fig:ctc}}\qquad
\caption{
\protect\subref{fig:mingap}
Simulation of blood flow through variable gap (in this case $6\mu m$) to obtain minimum gap size for each blood cell.
\protect\subref{fig:ctc}
Circulating tumor cell captured in magnetic bead trap (side and top view) from preliminary work \protect\cite{gusenbauer_self-organizing_2012}.
}
\label{fig:CTCsimulations}
\end{figure}

To obtain optimal channel gaps for differently blood cells the elastic properties of this cells are used and tested in a special simulation environment (Fig. \ref{fig:CTCsimulations}). The detailed description of the calibration process of the blood cells is described in preliminary work \cite{cimrak_modelling_2011}. The modeling of various tumor cells and other blood cells is ongoing work. The filter gap will be modified accordingly with the external magnetic gradient field.

We saw in experiments the bending of a membrane, but the displacement vs. the magnetic gradient force is not yet quantitatively measured. Simulations showed the influence of the a magnetic gradient field on the channel cross section of the flexible and tunable polymer. To get a much more detailed understanding of the fluid flow inside the channel further effort is needed. Gervais et. al \cite{gervais_flow-induced_2006} showed already the nontrivial fluidic behavior inside PDMS channels. The pressure drop that occurs in channels is much less than in rigid channels. Channel deformations decreases nonlinearly along the length of the channel. Also the aspect ratio (width to height) is important for the channel behavior. All the considerations done in their work need to be analyzed again with the influence of the magnetic active polymer.\\



\small{\textbf{Acknowledgment} The authors gratefully acknowledge the financial support
of Life Science Krems GmbH, the Research Association of Lower Austria.}\\

%

%

\begin{thebibliography}{}


\bibitem{lu_parylene_2010}
B. Lu, T. Xu, S. Zheng et. al, \textit{23rd International Conference on Micro Electro Mechanical Systems {(MEMS)}} ({IEEE}, 2010), 935-938

\bibitem{bell_isolation_2007}
D.W. Bell, D. Irimia, L. Ulkus et al., Nature \textbf{450}, 1235 (2007)

\bibitem{gertler_detection_2003}
R. Gertler, R. Rosenberg, K. Fuehrer et al., Recent Results in Cancer Research \textbf{162}, 149 (2003), {PMID:} 12790329

\bibitem{gusenbauer_self-organizing_2012}
M. Gusenbauer, A. Kovacs, F. Reichel et al., Journal of Magnetism and Magnetic Materials \textbf{324}, 977 (2012)

\bibitem{stott_isolation_2010}
S.L. Stott, C. Hsu, D.I. Tsukrov et al., Proceedings of the National Academy of Sciences \textbf{107}, 18392 (2010)

\bibitem{elmer}
CSC - IT Center for Science, accessed 16 July 2012, \texttt{http://www.csc.fi/english/pages/elmer}

\bibitem{abaqus}
Dassault Systemes, accessed 16 July 2012, \texttt{http://www.3ds.com/products/simulia/overview}

\bibitem{nanofer}
Nano Iron, accessed 16 July 2012, http://www.nanoiron.cz/en/nanofer-25


\bibitem{cimrak_modelling_2011}
I. Cimrak, M. Gusenbauer, T. Schrefl, Computers \& Mathematics with Applications \textbf{64}, 278-288 (2012)

\bibitem{gervais_flow-induced_2006}
T. Gervais, J. El-Ali, A. G\"unther et al., Lab on a chip \textbf{4}, 500-507 (2006)


\end{thebibliography}

\end{document}